\documentclass[
aps,
prl,
reprint,
superscriptaddress,
amsmath,
amssymb,
showpacs
]{revtex4-1}

\usepackage{graphicx}
\usepackage{float}
\usepackage{dcolumn}
\usepackage{bm}
\usepackage{amsmath}
\usepackage{amssymb}
\usepackage{latexsym}
\usepackage{epsfig}
\usepackage{amsbsy}
\usepackage{array}
\usepackage{amssymb}
\usepackage{setspace}
\usepackage{bm}
\usepackage{indentfirst}
\usepackage{multirow}
\usepackage{color}

\begin{document}


\title{Physical mechanism of the transverse instability in radiation pressure ion acceleration}


\author{Y. Wan}
\affiliation{Department of Engineering Physics, Tsinghua University, Beijing 100084, China}
\affiliation{Laser Fusion Research Center, China Academy of Engineering Physics, Mianyang, Sichuan 621900, China}
\author{C. -H. Pai}
\affiliation{Department of Engineering Physics, Tsinghua University, Beijing 100084, China}
\author{C. J. Zhang}
\author{F. Li}
\affiliation{Department of Engineering Physics, Tsinghua University, Beijing 100084, China}
\affiliation{Laser Fusion Research Center, China Academy of Engineering Physics, Mianyang, Sichuan 621900, China}
\author{Y. P. Wu}
\author{J. F. Hua}
\author{W. Lu}
\email[]{weilu@tsinghua.edu.cn}
\affiliation{Department of Engineering Physics, Tsinghua University, Beijing 100084, China}
\author{Y. Q. Gu}
\affiliation{Laser Fusion Research Center, China Academy of Engineering Physics, Mianyang, Sichuan 621900, China}
\author{L. O. Silva}
\affiliation{Fus$\tilde{a}$o Nuclear¡ªLaborat$\acute{o}$rio Associado, Instituto Superior Tecnico, 1049-001 Lisbon, Portugal}
\author{C. Joshi}
\author{W. B. Mori}
\affiliation{University of California Los Angeles, Los Angeles, CA 90095, USA}

\date{\today}

\begin{abstract}
The transverse stability of the target is crucial for obtaining high quality ion beams using the laser radiation pressure acceleration (RPA) mechanism. In this letter, a theoretical model and supporting two-dimensional (2D) Particle-in-Cell (PIC) simulations are presented to clarify the physical mechanism of the transverse instability observed in the RPA process. It is shown that the density ripples of the target foil are mainly induced by the coupling between the transverse oscillating electrons and the quasi-static ions, a mechanism similar to the transverse two stream instability in the inertial confinement fusion (ICF) research. The predictions of the mode structure and the growth rates from the theory
agree well with the results obtained from the PIC simulations in various regimes, indicating the model contains the essence of the underlying physics of the transverse break-up of the target.
\end{abstract}


\maketitle

Recently, laser radiation pressure ion acceleration (RPA) has attracted much attention due to its great potential for building very compact ion accelerators that can be used in diverse fields such as medical therapy\cite{Bulanov2002, Linz2007}, ion radiography\cite{Borghesi2002}, generation of short-lived isotopes needed in positron emission tomography\cite{Spencer2001}, injectors for conventional accelerators\cite{Krushelnick2000}, fast ignition fusion research\cite{Roth2001} and so on.
Ideal one dimensional (1D) simulations show monoenergetic ion acceleration in the RPA process using a circularly polarized (CP) laser pulse\cite{Macchi2005,Xiaomei2007,Yan2008,klimo2008,Robinson2008,Macchi2013,Daido2012} with high energy conversion efficiency.
In reality, however, the finite transverse witdth of the laser pulse can deform the target shape, leading to electron heating and energy spectrum broading of the accelerated ions\cite{klimo2008,Robinson2008,Chen2008}.
At the same time, 2/3D simulations also show that transverse density ripples can grow significantly, leading to some of the laser energy through and breaking up the target\cite{klimo2008,Robinson2008,Chen2008,Chen2009,Yan2009,Qiao2011,Xiaomei2011,Wu2014}. This phenomenon shows up even for a laser pulse of infinite width and uniform intensity profile \cite{Pegoraro2007,Robinson2008,Wu2014}.
Various mechanisms have been proposed to explain the structrue of these ripples, such as Rayleigh-Taylor like (RT-like) instability\cite{Pegoraro2007,klimo2008,Robinson2008,Palmer2012,Qiao2011,Wu2014,Eliasson2015,Sgattoni2015}, Weibel like instability\cite{Yan2009,Xiaomei2011} and so on.
However, these models have not been able to give accurate predictions of the mode structure and its growth rates for a wide range of laser and plasma parameters.

\begin{figure}[htbp]
\centering
 \includegraphics[width=0.45\textwidth]{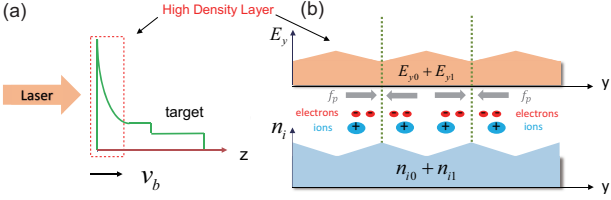}%
 \caption{\label{physical_picture}(a) The schematic model of hole boring process by radiation pressure. (b) The physical picture of transverse instability within the high density layer. The z and y axis represent the longitudinal and transverse directions, respectively. The $n_{i1}$ and $E_{y1}$ represent the ion density and the transverse electric field fluctuations, respectively. $f_p$ represents the ponderomotive force.}
\end{figure}

In this letter, we show through theoretical analysis and PIC simulations that these surface ripples are more likely induced by the coupling between the transverse oscillating electrons and the quasi-static ions within the high density layer formed by the laser radiation pressure pushing the surface plasma forward in a process often called 'hole-boring"  (H-B)\cite{Macchi2005,Robinson2009}. As shown in Fig.~\ref{physical_picture}(a), during this H-B process, soon after the laser impinges on the front surface of the target, a dynamic equilibrium between the laser pressure and the electrostatic field within the plasma is built, forming a quasi-static high density structure co-moving with the laser pulse\cite{Macchi2005}. Within this layer, the CP laser field oscillates at the laser frequency along both transverse directions albeit pi radians out of phase. A very small transverse ion density fluctuation can couple with the oscillating laser field to excite an electron oscillation. This oscillation in turn can couple with the oscillating laser field to generate a ponderomotive force with spatial variation, driving the electrons to enhance the ion density fluctuation. The physical picture of this process is illustrated in Fig.~\ref{physical_picture}(b).
It is indeed very similar to the oscillating two stream instability extensively studied in the inertial confinement fusion (ICF) research \cite{Silin1965,Dubois1965,Sanmarti1970}.
However, there are significant differences. First, the oscillating laser field only exists within the narrow layer formed by the laser pressure, and its amplitude is determined by the boundary conditions at the interface. And second, in the case of RPA, the laser is relativistically intense, with the normalized vector potential $a_0$ on the order of 1, much larger than those studied in the previously considered oscillating two stream instability.


We first derive a 1D theoretical model of this instability based on the above physical picture, and then verify it using 2D PIC simulations. For simplicity, a relativistic cold two-fluid plasma description is adopted, and only electrostatic perturbations along the laser electric field are considered.

In the co-moving frame of the high density layer, the cold fluid equations for electrons and protons in the transverse direction are:
\begin{subequations}
\begin{eqnarray}
  \frac{\partial n_{(i,e)}}{\partial t} +\frac{\partial n_{(i,e)} v_{(i,e)y}}{\partial y}=0 \label{flu_con}\\
  \frac{\partial P_{(i,e)y}}{\partial t}+v_{(i,e)y}\frac{\partial P_{(i,e)y}}{\partial y}=q_{(i,e)} E_y\label{flu_mon}\\
  \frac{\partial E_{y}}{\partial y}=4\pi e(n_{i}-n_{e})\label{flu_pos}
\end{eqnarray}
\end{subequations}
where y is the transverse direction; v and P are the velocity and momentum, respectively. $v_{(i,e)y}=P_{(i,e)y}/(1+P_{(i,e)x}^2+P_{(i,e)y}^2)^{1/2}$. For simplicity, we assume the fluctuation only depends on (y, t) and ions are non-relativistic.
To linearize the fluid equations, all the quantities can be decomposed as a stationary part plus a first order quantity, such as $v_{ey}=v_{e0}+v_{e1}, P_{ey}=P_{e0}+P_{e1},v_{iy}=v_{i0}+v_{i1} (v_{i0}=0), n_e=n_0+n_{e1}, n_i=n_0+n_{i1}, E_y=E_{y0}+E_{y1},
$ where $E_{y0}=E_0\cos{(\omega_0t+\phi)}$, and $P_{e0}=P_{os}\sin{(\omega_0t+\phi)}$.
By using the standard Fourier analysis (assuming all first order quantities have the form of $\exp(iky-wt)$), one can get the following equations after eliminating $n_{i1},v_{i1},E_{y1}$:
\\
\begin{subequations}
\begin{eqnarray}
  -i\omega n_{e1}(\omega)-\frac{v_{os} k}{2}[n_{e1}(\omega+\omega_0)\nonumber\\
  -n_{e1}(\omega-\omega_0)]+i k n_{0}\kappa P_{e1}(\omega)=0 \\
 -i\omega P_{e1}(\omega)-\frac{v_{os} k}{2}[P_{e1}(\omega+\omega_0)\nonumber\\
 -P_{e1}(\omega-\omega_0)]-\epsilon(\omega)n_{e1}(\omega)=0
 \end{eqnarray}
 \end{subequations}
 where $v_{os}=\frac{P_{os}}{\gamma_0}$ is the electron quiver velocity amplitude in the laser electric field, $\gamma_0$ is the electron's zero-order relativistic factor. $\omega_0$ and $\omega_{pi}$ are the laser frequency and ion plasma frequency respectively, and $\epsilon(\omega)=-i\frac{4\pi }{k}\frac{\omega^2}{\omega_{pi}^2-\omega^2}, \kappa=\frac{2-v_{os}^2}{2\gamma_0}$.

These two equations show the relationship between $n_{e1}$ and $P_{e1}$ at $\omega$ and $\omega\pm\omega_0$. By replacing $\omega$ with $\omega\pm\omega_0$, one can obtain six equations describing the relationship among $\omega$, $\omega\pm\omega_0$ and $\omega\pm 2\omega_0$.
However, to obtain a close dispersion relation, further assumption is needed. Since the dynamics involves ion density evolution, which is typically on a much slower time scale than the laser oscillation, we may drop all the fast time scale terms at $\omega\pm 2\omega_0$.
Therefore, we now have six equations for six quantities ($n_{e1}$, $P_{e1}$ at $\omega$ and $\omega\pm\omega_0$), and this can be casted into a matrix form as follows:
 \begin{widetext}
 \begin{equation}\label{matrix}
\left(
\begin{array}{cccccc}
-i\omega&-v_{os}k/2&v_{os}k/2& ikn_0\kappa&0&0\\
v_{os}k/2&-i(\omega+\omega_0)&0&0&ikn_0\kappa&\\
-v_{os}k/2&0&-i(\omega-\omega_0)&0&0&ikn_0\kappa\\
\epsilon(\omega)&0&0&-i\omega&-v_{os}k/2&v_{os}k/2\\
0&i\frac{4\pi }{k}&0&v_{os}k/2&-i(\omega+\omega_0)&0 \\
0&0&i\frac{4\pi }{k}&-v_{os}k/2&0&-i(\omega-\omega_0) \\
\end{array}
\right)
\left(
\begin{array}{c}
n_{e1}(\omega)\\
n_{e1}(\omega+\omega_0)\\
n_{e1}(\omega-\omega_0)\\
P_{e1}(\omega)\\
P_{e1}(\omega+\omega_0)\\
P_{e1}(\omega-\omega_0)\\
\end{array}
\right)
=\vec{0}
\end{equation}
\end{widetext}
The dispersion relation can be obtained by taking the determinant of the matrix equal to zero. To get the growth rate, we solve the dispersion equation for each real k value, and obtain the imaginary part of $\omega$ (Im($\omega$)).
The wave number for the mode with the maximal growth rate ($k_m$) can be calculated numerically by taking the maximal value of $|Im(\omega)|$.

\begin{figure}[H]
\centering
 \includegraphics[width=0.3\textwidth]{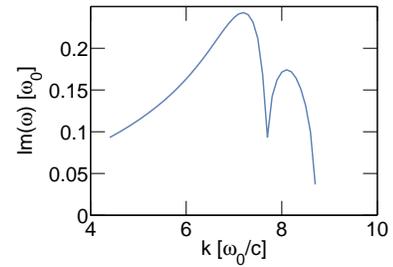}%
 \caption{\label{matrix_example} The relationship between k and Im($\omega$) for the case of $\gamma_{0}=1.5$, $\omega_{pe}=6\ \omega_0$, $\omega_{pi}=0.13\ \omega_0$. }
\end{figure}

Fig.~\ref{matrix_example} shows an example. We take $\gamma_{0}=1.5$, $\omega_{pe}=6\ \omega_0$, $\omega_{pi}=0.13\ \omega_0$, and $v_{os}=\sqrt{1-1/\gamma_0^2}$. The relation between k and Im($\omega$) is presented in Fig.\ref{matrix_example} and $k_m$ = 7.2 $\omega_0/c$.

The dispersion relation can be simplified significantly for $a_0<1$ by taking $\kappa\approx 1$ for non-relativistic electrons and also keeping the dominant terms:
\begin{eqnarray}
\omega^4(\xi^4+12\omega_{pe}^4)-\omega^2\omega_{pe}^2(\xi^2-2\omega_{pe}^2)^2
\nonumber\\
+\xi^2\omega_{pe}^2\omega_{pi}^2(\xi^2-2\omega_{pe}^2+2\omega_0^2)=0
\label{dispersion}
\end{eqnarray}
where $\xi=k v_{os}$ and $\omega_{pe}$, $\omega_{pi}$ are the electron and ion plasma frequencies of the high density layer at the foil's front.  $k_m$ can be directly solved from Eq.\ref{dispersion}:
\begin{eqnarray}
k_m v_{os}\approx\sqrt{2}
\omega_{pe}\label{kv}
\end{eqnarray}
A simple estimation of $\omega_{pe}$ can be obtained for $a_0<1$ by assuming an uniform density profile and charge neutrality (i.e. $n_e\approx n_i$) within the high density layer. Then the electrostatic field $E_s$ can be described as $E_{s}=E_{s0}(l_s-z)/l_s,\ (0\leq z\leq l_s)$, where $E_{s0}$ is the maximum longitudinal electrostatic field, and $l_s$ is the thickness of this layer. In the hole boring process, after balance is built, the equilibrium between the electrostatic force and the radiation pressure within the layer can be written as
$\frac{1}{2}E_{s0}el_sn_e=\frac{2I}{c}$\cite{klimo2008}, where $n_e$ is the averaged ion density within the layer. In the co-moving frame, ions are moving into this area with $v_b$ and satisfy $\frac{1}{2}m_iv_b^2=\frac{1}{2}E_{s0}l_s$, where $v_b$ is the hole boring velocity. Meanwhile ions are also moving out of this layer with a velocity of $v_b$. Therefore, during $\delta t$, the momentum conservation relation leads to $m_in_{p0}v_b\delta t(2v_b)=\frac{2I}{c}\delta t$\cite{Macchi2005}, where $n_{p0}$ is the initial plasma density. Combining these three equations, we get $n_e=n_i=4n_{p0}$. This simple relation can be readily verified by PIC simulations.

On the other hand, by applying the Fresnel-like boundary condition and neglecting the $v_b\times B$ effect in the $y$ direction ($v_b\ll c$), we get $v_{os}/c\approx 2a_0\frac{\omega_0}{\omega_{pe}}$.
With the new form of $v_{os}$ and $\omega_{pe}$, Eq.~\ref{kv} can be written in a form easier for direct comparison with PIC simulations:
\begin{eqnarray}
  k_m\approx 2\sqrt{2}\frac{n_{p0}}{a_0n_c}[\omega_0/c]\label{km_expression}
\end{eqnarray}
where $n_c=\frac{m_e\omega_0^2}{4\pi e^2}$ is the critical density.
One can see that $k_m$ has a very simple dependence on $a_0$ and $n_{p0}$.

To verify the above theory, we performed a series of 2D PIC simulations using the code OSIRIS\cite{Fonseca2002}. In these simulations, a CP laser driver with a transverse uniform profile is used, and the laser propagates in the z direction.
High resolutions are used in both directions ($\Delta y=\Delta z= 0.002\ c\omega_0^{-1}$), with 16 particles in each cell.
The foil is a pure hydrogen plasma with a step density profile.

\begin{figure}[H]
\centering
 \includegraphics[width=0.45\textwidth]{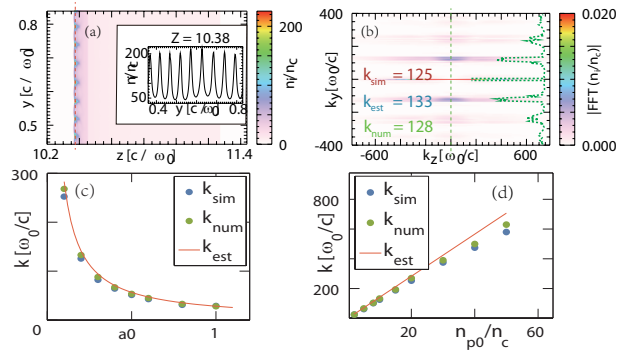}%
 \caption{\label{a=0.2_n=10_ka0n0_relation}(a) In the case of $a_0=0.2$, $n_{p0}=10n_c$, the proton density with ripples in the front high density layer and its lineout distribution at $z=10.38\ c/\omega_0$ (the red dot line). (b) the FFT of the proton density and its lineout distribution at $k_z=0$ (the green dot line). (c) The relationship between $k_m$ and $a_0$ when $n_{p0}=10n_c$. (d) The relationship between $k_m$ and $n_{p0}$ when $a_{0}=0.2$.
 $k_{sim}$, $k_{est}$ and $k_{num}$ are obtained from PIC simulations, from direct numerical solutions of Eq.~\ref{matrix}, and from Eq.~\ref{km_expression}, respectively.}
\end{figure}

 Fig.~\ref{a=0.2_n=10_ka0n0_relation} (a) and (b) show an example. We begin with $a_0=0.2$ and $n_{p0}=10n_c$. In Fig.~\ref{a=0.2_n=10_ka0n0_relation} (a), one can see density ripples are induced in the high density layer irradiated by the laser pulse. A lineout corresponding with the red dot line of Fig.~\ref{a=0.2_n=10_ka0n0_relation} (a) is presented, showing the periodic density structures appearing during the interaction process. Fig.~\ref{a=0.2_n=10_ka0n0_relation} (b) is the 2D Fourier Transformation of Fig.~\ref{a=0.2_n=10_ka0n0_relation} (a) and a lineout showing the distribution of $k_y$ at $k_z=0$ corresponding with the green dot line is also presented. It clearly indicates that the instability mode number $k_m$ is about 125 $\omega_0/c$, which have good agreements with the estimated value $k_{est}$ (133 $\omega_0/c$) from Eq.~\ref{km_expression} and numerical value $k_{num}$ (128 $\omega_0/c$) from Eq.~\ref{matrix}.
In Fig.\ref{a=0.2_n=10_ka0n0_relation}(c), we plot the relation between $k_m$ and $a_0$ by fixing the plasma density ($n_{p0}=10n_c$). Three values of $k_m$ ($k_m$ obtained from PIC simulation, from direct numerical solution of Eq.~\ref{matrix}, and from Eq.\ref{km_expression}) are used for comparison. One can see that a very good agreement is obtained. In Fig.\ref{a=0.2_n=10_ka0n0_relation}(d), we also plot the relation between $k_m$ and $n_{p0}$ by fixing $a_0=0.2$. One can see equally good agreements between the three values of $k_m$.

Eq.~\ref{dispersion} can also give a simple expression of the growth rate $\gamma_{m0}$ at $k_m$.
\begin{eqnarray}
\gamma_{m0}\approx 2\ \omega_{pi}\label{rm0}
\end{eqnarray}

We performed a series of 2D simulations with a large range of plasma parameters similar to Fig.~\ref{a=0.2_n=10_ka0n0_relation} to confirm our analysis of the growth rates.

\begin{figure}[H]
\centering
 \includegraphics[width=0.48\textwidth]{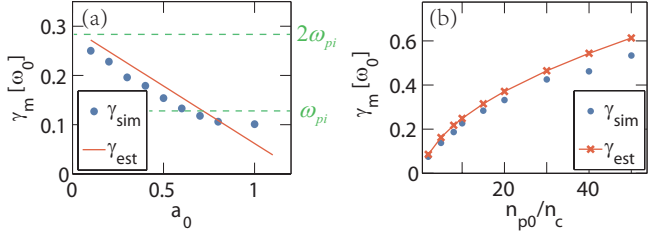}%
 \caption{\label{ra0n0_relation}(a) The relationship between $\gamma_m$ and $a_0$ at $n_{p0}=10n_c$. (b) The relationship between $\gamma_m$ and $n_{p0}$ at $a_{0}=0.2$.$\gamma_{est}$ and $\gamma_{sim}$ are the estimated and simulation wave numbers of ion density ripples respectively}
\end{figure}

Fig.~\ref{ra0n0_relation}(a) shows the relation between $\gamma_m$ and $a_0$ at $n_{p0}=10n_c$. It is found that though $\gamma_m$ is varying with $a_0$, it is still on the same order of $\omega_{pi}$ (in the range of $\omega_{pi}\sim2\omega_{pi}$), which has some agreements with Eq.~\ref{rm0}. The weak relation between $\gamma_m$ and $a_0$ mainly comes from the fact that in the co-moving frame, protons are moving in and out of the high density layer consecutively, and this area is not stationary. If the longitudinal flow is quite slow, the expression of growth rate $\gamma_{m}$ can also be evaluated. We assume that at $t=t_0$, the ion density fluctuation is $f(t_0)=\delta n_{0}l_s$, where $l_s$ is the length of the high density layer, and $\delta n_0$ is the ion density fluctuation at $t=t_0$.
Then at $t=t_0+\delta t$, the fluctuation becomes as $f(t_0+\delta t)=\delta n_0 e^{\gamma_{m0}\delta t}(l_s-v_b\delta t)$. The growth rate can be calculated as $e^{\gamma_{m}(t_0)\delta t}=f(t_0+\delta t)/f(t_0)$, where $v_b$ is the hole boring velocity of ions moving in or out of this region. Based on the analysis above, it is straightforward to obtain:
\begin{eqnarray}
\gamma_{m}\approx 2\omega_{pi}-2\omega_{0i}\eta a_0\label{rmt}
\end{eqnarray}
where $\omega_{0i}=\sqrt{m_e/M_i}\omega_0$ is the critical ion plasma frequency, and $\eta$ is a coefficient. Eq.~\ref{rmt} shows $\gamma_{m}$ has a weakly linear dependence on $a_0$, which has quite good agreement with Fig.\ref{ra0n0_relation}(a) for $a_0<0.7$. And $\eta\approx 4.8$ can be evaluated from simulations. Eq.~\ref{rmt} is valid for the initial several $1/\omega_{pi}$, since as the instability grows, more other effects like electron heating and radiation pressure transverse nonuniformity will get involved.

As Fig.~\ref{ra0n0_relation}(b) shows, if we fix $a_0=0.2$, the values of growth rates from simulations also have good agreements with that from Eq.\ref{rmt}.

For $a_0>1$, the relativistic factor of electrons need to be considered. Similar to Eq.~\ref{km_expression}, a simple expression of $k_m$ can also be approximately obtained as:
\begin{eqnarray}
  k_m\approx\sqrt{2}\frac{\omega_{pe}}{v_{os}}\sqrt{\kappa}=
  \frac{\omega_{pe}}{\sqrt{\gamma_0}}\sqrt{
  \frac{\gamma_0^2+1}{\gamma_0^2-1}}
\label{km_relativistic}
\end{eqnarray}
where $\gamma_0$ is electron's zero-order quiver energy.
Eq.~\ref{km_relativistic} is valid both for thick foil cases (Hole boring) \cite{Macchi2005,Robinson2009}and thin foil cases (Light sail)\cite{Xiaomei2007,klimo2008,Yan2008,Robinson2008,Qiao2009,Macchi2009}. However,
the expressions of $\omega_{pe}$ and $\gamma_0$ can not be easily obtained directly. Instead, we use the estimations from simulations.
\begin{figure}[H]
\centering
 \includegraphics[width=0.5\textwidth]{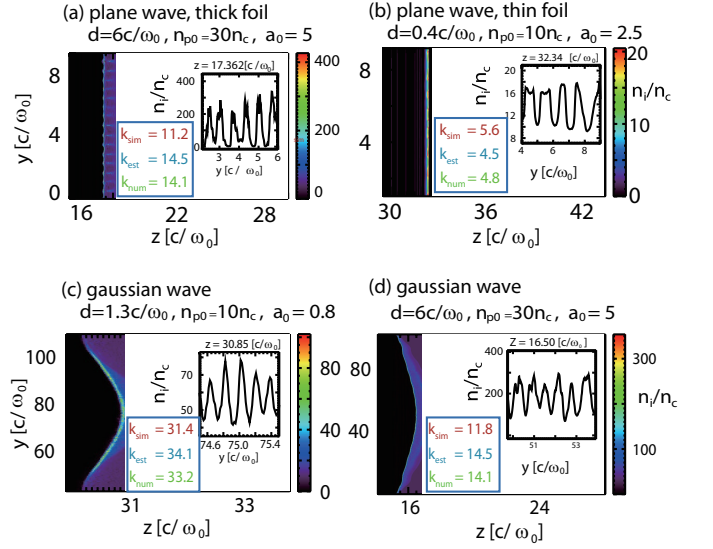}%
 \caption{\label{4examples}(a)-(b) for $a_0>1$, proton densities of two different scenarios (thick foil case (a) and thin foil case (b))are presented. (c)-(d) laser pulses with transverse Gaussian profile are used to interact with foils to confirm the usability of the theoretical expressions.  $k_{sim}$, $k_{est}$ and $k_{num}$ are obtained from PIC simulations, from direct numerical solutions of Eq.~\ref{matrix}, and from Eq.~\ref{km_relativistic}, respectively. d represents the thickness of the target.}
\end{figure}
To check the validity of Eq.~\ref{km_relativistic} for $a_0>1$, we performed 2D PIC simulations for two different scenarios (thick foil case and thin foil case) and plot the typical results in Fig.~\ref{4examples}(a) and (b). In Fig.~\ref{4examples}(a), a circularly polarized laser ($a_0=5$) is used to interact with a thick target(thickness $d=6\ c/\omega_0$, initial density $30\ n_c$). The mode wave number in the simulation is $11.2\ \omega_0/c$, which is similar to the estimated value $14.5\ \omega_0/c$ from Eq.~\ref{km_relativistic} and numerical value $14.1\ \omega_0/c$ from Eq.\ref{matrix}.
In Fig.~\ref{4examples}(b), a circularly polarized laser ($a_0=2.5$) is used to interact with a thin target(thickness $d=0.4\ c/\omega_0$, initial density $10\ n_c$). The mode wave number in the simulation is $5.6\ \omega_0/c$, which is similar to the estimated value $4.5\ \omega_0/c$ from Eq.~\ref{km_relativistic} and numerical value $4.8\ \omega_0/c$ from Eq.\ref{matrix}.

In all the above simulations, uniform laser intensity profiles are used for the exact comparison with the theoretical model. In more realistic cases, the laser typically has nonuniform intensity profiles like Gaussian( e.g., $\exp(-r^2/w_0^2)$). To confirm the usability of the theoretical expressions (Eq.~\ref{matrix} and \ref{km_relativistic}), we also performed simulations and plot the typical results in Fig.~\ref{4examples}(c) and (d) .

For $a_0<1$, in Fig.~\ref{4examples}(c), a CP laser pulse with $a_0=0.8$ and a radius \(w_0=20\ c/\omega_0\) is used to interact with a thin foil ( thickness $1.3\ c/\omega_0$, initial density $n_{p0}=10n_c$). The mode wave number in the simulation is $31.4\ \omega_0/c$, which is similar to the estimated value $34.1\ \omega_0/c$ from Eq.~\ref{km_relativistic} and numerical value $33.2\ \omega_0/c$ from Eq.\ref{matrix}. For $a_0>1$, in Fig.~\ref{4examples}(d), a CP laser pulse with $a_0=5$ and a radius \(w_0=40\ c/\omega_0\) is used to interact with a thick foil ( thickness $6\ c/\omega_0$, initial density $n_{p0}=30n_c$). The mode wave number in the simulation is $11.8\ \omega_0/c$, which is similar to the estimated value $14.5\ \omega_0/c$ from Eq.~\ref{km_relativistic} and numerical value $14.1\ \omega_0/c$ from Eq.\ref{matrix}.


In conclusion, we have demonstrated that the surface ripples in the RPA process are mainly induced by the coupling between fast oscillating electrons and quasi-static ions within the high density layer formed by the laser pressure. A one-dimensional model is presented here to predict the mode structure and its growth rate, which has good agreement with 2D PIC simulation results.

\begin{acknowledgments}
This work was supported by the National Basic Research
Program of China No. 2013CBA01501, NSFC Grant No.
11425521, No. 11535006, No. 11175102, No.11005063,
No. 11375006 and No. 11475101, the Foundation of CAEP
No. 2014A0102003, Tsinghua University Initiative Scientific Research Program, the Thousand Young Talents Program, DOE Grants No. DE-FG02-92-ER40727, No. DE-SC0008491, and No. DE-SC0008316, and NSF Grants No. PHY-0936266, No. PHY-0960344, and No. ACI-1339893. Simulations are performed on Hoffman cluster at UCLA and Hopper cluster at National Energy Research Scientific Computing Center.
(NERSC).
\end{acknowledgments}

\bibliography{transverse_instability_ref}

\end{document}